\documentclass[aip,reprint,amsmath,amssymb,nofootinbib,superscriptaddress]{revtex4-2}
\usepackage{graphicx}
\usepackage{dcolumn}
\usepackage{bm}
\usepackage{amsmath}
\usepackage{epstopdf}
\usepackage{color}
\usepackage{verbatim}
\usepackage{caption}
\usepackage{subfig}
\usepackage{float}
\usepackage{appendix}
\usepackage{ulem}

\begin{document}
\title{Mechanical response of quasi-two-dimensional colloidal clusters under uniaxial tension}
\author{Yanhui Yang$^{\dag}$\footnotemark[1]}
\affiliation{Department of Physics and Center for Complex Flow and Soft Matter Research, Southern University of Science and Technology, Shenzhen, Guangdong, China}
\author{Jiawei Kang$^{\dag}$\footnotemark[1]}
\affiliation{School of Physics and Key Laboratory of Functional Polymer Materials of Ministry of Education, Nankai University, and Collaborative Innovation Center of Chemical Science and Engineering, Tianjin, China}
\author{Yao Li}
\email{liyao@nankai.edu.cn}
\affiliation{School of Physics and Key Laboratory of Functional Polymer Materials of Ministry of Education, Nankai University, and Collaborative Innovation Center of Chemical Science and Engineering, Tianjin, China}
\author{Xiaoguang Ma}
\email{maxg@sustech.edu.cn}
\affiliation{Department of Physics and Center for Complex Flow and Soft Matter Research, Southern University of Science and Technology, Shenzhen, Guangdong, China}
\renewcommand\thefootnote{}
\footnotetext[1]{$^\dag$ These authors contributed equally.}
\date{\today}
	
\begin{abstract}
Despite extensive studies of equilibrium conformations of colloidal clusters, little is known about their mechanical response. Here, we investigate the tensile behavior of a quasi-two-dimensional colloidal cluster subjected to uniaxial tension up to fracture. The sample is a ribbon-shaped assembly of 16 colloidal beads bound by short-range depletion attraction. Using multiple optical tweezers, we clamp the cluster at both ends and perform a tensile test along its long axis. Combining video microscopy with particle tracking, we measure the tensile stress, strain, and particle configurations during deformation. We observe diverse mechanical response behaviors, including elastic, plastic, and soft-mode deformation, with fracture occurring at a strain near 10\%. To explain these behaviors, we construct a spring-mass frame model with breakable elastic bonds. We perform canonical Monte Carlo simulations on the full model with 32 degrees of freedom and compute the statistical distributions of mechanical observables using a simplified model with only 7 degrees of freedom. Both the simulations and the theoretical calculations accurately reproduce the experimental stress--strain curves. Moreover, the configuration distributions predicted by the simplified model agree well with both experiment and simulation in the elastic and soft-mode regimes, with only minor discrepancies in the plastic regime. This work demonstrates that the simplified spring-mass model captures the essential physics governing the rich tensile response behavior of the colloidal cluster.
\end{abstract}
\maketitle

\section{Introduction}
\label{sec:introduction}
Clusters composed of a small number of atoms and particles are mesoscopic building blocks of various disordered particulate matter, spanning liquids, gels, polymers, glasses, and biological condensates \cite{BERNAL1960,Groenewold2001,ALEXANDER199865,Mossa2004,Sciortino2005,annurev:/content/journals/10.1146/annurev-conmatphys-031016-025357,PhysRevLett.94.208301,doi:10.1126/science.aaf4382,PhysRevE.85.051403,10.1063/1.4774076,PhysRevE.91.012303,D3SM01717F,D4SM00811A,kmpn-zjk7}. Local packing, structural order, growth, and configurational transformations of clusters strongly influence the thermodynamic and mechanical properties of their assemblies at the macroscopic scale \cite{PhysRevLett.106.225503,2019-Ma-10.1063/1.5091564,Rouwhorst2020,doi:10.1073/pnas.1206742109,doi:10.1021/acs.jpcb.4c06392,doi:10.1126/sciadv.aav6090,Whitaker2019,Bantawa2023,doi:10.1126/sciadv.abb8107,Tsurusawa2023,RevModPhys.96.045003}.

Colloidal particles assembled via short-range attractive forces serve as convenient hard-sphere model systems for studying molecular and atomic clusters. Despite the differences in length scales, colloidal and atomic clusters both exhibit characteristic polyhedral and icosahedral symmetries, making colloids a convenient experimental platform for probing structural order and rearrangement dynamics that are difficult to access directly in their atomic counterparts \cite{Groenewold2001,Mossa2004,PhysRevLett.94.208301,Sciortino2005,doi:10.1126/science.1086189,PhysRevE.91.012303,annurev:/content/journals/10.1146/annurev-conmatphys-031016-025357,RevModPhys.96.045003}. In addition, synthetic colloids and liquid droplets can be engineered to possess tunable size, shape anisotropy, surface chemistry, competing interactions, and nonequilibrium properties \cite{doi:10.1126/science.1086189,doi:10.1126/science.1197451,Aubret2021,doi:10.1126/science.aaf4382,PhysRevLett.121.138002,Lim2019,doi:10.1073/pnas.2001272117,McMullen2022,rjk2-q2wh,Hooshanginejad2024}. This versatility enables the bottom-up design and hierarchical fabrication of functional particulate materials constructed from customizable structural motifs \cite{doi:10.1126/science.1086189,2019-Cao-Orientational-NP}. Harnessing these technological opportunities requires a comprehensive understanding of the structure, mechanics, and dynamics of these clusters under both equilibrium and nonequilibrium conditions.

Current understanding of the equilibrium thermodynamic behavior of colloidal clusters has focused mainly on their ground-state conformations. Under the ``sticky-sphere'' approximation, interparticle attractions are represented by a ``sticky'' parameter determined solely by pairwise interactions and assumed to be independent of aggregate geometry and structural rearrangement \cite{PhysRevE.105.014117,PhysRevLett.103.118303,doi:10.1126/science.1181263,PhysRevLett.105.068001,doi:10.1137/100784424,perry2012real,doi:10.1073/pnas.1211720110,PhysRevLett.114.228301,doi:10.1137/140982337,PhysRevE.95.022130,annurev:/content/journals/10.1146/annurev-conmatphys-031016-025357,PhysRevE.98.032608,PhysRevE.105.014117}. As a result, the conformational problem reduces to enumerating distinct geometric states that follow specific optimization principles \cite{doi:10.1137/100784424,doi:10.1137/140982337}. For unconstrained clusters, for example, the number of interparticle contacts is maximized to minimize potential energy \cite{PhysRevE.85.051403,doi:10.1073/pnas.1211720110,PhysRevLett.114.228301}. In contrast, packings formed under capillary confinement are known to minimize the second moment of the mass distribution \cite{doi:10.1126/science.1086189}.

Despite this success in characterizing equilibrium cluster conformations, much less is known regarding how small particle assemblies respond to external loading \cite{PhysRevMaterials.8.035604,rjk2-q2wh}. This phenomenon exists even in bulk disordered matter under zero external stress, where {\it internal} stress is typically localized and involves only a subset of load-bearing particles that form clusters. Characterizing deformation behavior at the cluster scale, both experimentally and theoretically, remains largely unexplored. Recent studies demonstrate that the full interparticle potential energy, instead of simple contact counting, is necessary for accurate predictions of the tensile and yield properties of polymer chains \cite{PhysRevLett.134.218101}. Yet it remains an open question whether incorporating detailed interparticle potential information can faithfully predict rheological and mechanical properties of attractive colloids \cite{ALEXANDER199865,PhysRevE.70.040401,PhysRevLett.103.208301,doi:10.1126/sciadv.aav6090,Whitaker2019,Bantawa2023}.

In this paper, we combine experiment, simulation, and theory to investigate the tensile behavior of quasi-two-dimensional (quasi-2D) colloidal clusters bound by depletion forces (see Fig.~\ref{fig:Fig1}). Our sample consists of 16 micron-sized colloidal beads assembled into a ``ribbon'' geometry, one of the simplest cluster configurations that exhibits nontrivial mechanical responses under loading. The ground-state structure of the cluster is a zigzag chain of close-packed particles forming equilateral triangles. These triangles are the 2D analogues of tetrahedra in three-dimensional (3D) particulate systems, making this study relevant to 3D clusters as well\cite{BERNAL1960,PhysRevLett.94.208301}. Using multiple optical tweezers, we apply quasistatic uniaxial tension along the cluster's long axis. Via video microscopy and single-particle tracking, we measure tensile stress, tensile strain, and evolving particle configurations until fracture. The resulting stress--strain curve exhibits sequential deformation stages, including multiple elastic and plastic regimes as well as a low-energy, soft-mode-like deformation, with a fracture strain of approximately $10\%$.

To explain these observations, we develop a deformable and breakable spring-network framework to describe interparticle bonds within the cluster. This full model is employed for canonical Monte Carlo simulations. A major goal of this work is to identify the dominant parameters governing the cluster's response. To this end, we reduce the total number of degrees of freedom from 32 to 7 to simplify the full sping framework. This dimensionality reduction enables us to compute the partition function of the system in a seven-dimensional configuration space, from which we obtain ensemble-averaged stress, strain, and configurational distributions. Both simulation and theoretical predictions quantatively reproduce the experimentally observed elastic, plastic, and soft-mode regimes, confirming that the simplified model captures the key characteristics of the tensile behavior of the cluster. Comparisons of configurational distributions reveal good quantitative agreement among experiment, simulation, and theory in the elastic and soft-mode regimes, while discrepancies emerge only in the plastic regime, thereby showing the limitations of the reduced model. Overall, our findings elucidate the deformation features of small clusters with simple ground-state geometries. The underlying mechanism may provide insights for designing colloidal materials with tunable mechanical performance.

The rest of this paper is organized as follows. Section \ref{sec:experiment} describes sample preparation and mechanical testing procedures. Section \ref{sec:theory} details the spring-network framework used for Monte Carlo simulations and the reduced model for analytical predictions. Section \ref{sec:results} presents the experimental data and compares them with theoretical and numerical results. Finally, Sec.~\ref{sec:summary} summarizes the work.

\section{Experiment}
\label{sec:experiment}
The colloidal cluster comprises monodisperse polystyrene microspheres with a nominal diameter $D=1.36\pm 0.02$ $\mu$m (Thermo Scientific). Prior to use, the particles are thoroughly washed via repeated sonication and centrifugation in deionized water to remove chemical contaminants and particle aggregates. The cleaned particles are resuspended in an aqueous solution containing $60$ mM hexaethylene glycol monododecyl ether ($\text{C}_{12}\text{E}_6$) and $1$ mM sodium chloride (NaCl). The $\text{C}_{12}\text{E}_{6}$ micelles are employed as temperature-tunable depletants to provide short-ranged attractions for the colloidal particles \cite{2021-Ma-10.1063/5.0059084,2023-Ma-10.1063/5.0146155}. At $39\pm 0.1^\circ\text{C}$, the interparticle attraction strength---the minimal interparticle potential---reaches approximately $-6.5$ $k_\text{B}T$ at an interparticle separation $d\simeq 1.41$ $\mu$m (see Appendix \ref{appendix_a}).
\begin{figure}[H]
	\centering\includegraphics[width=1\linewidth]{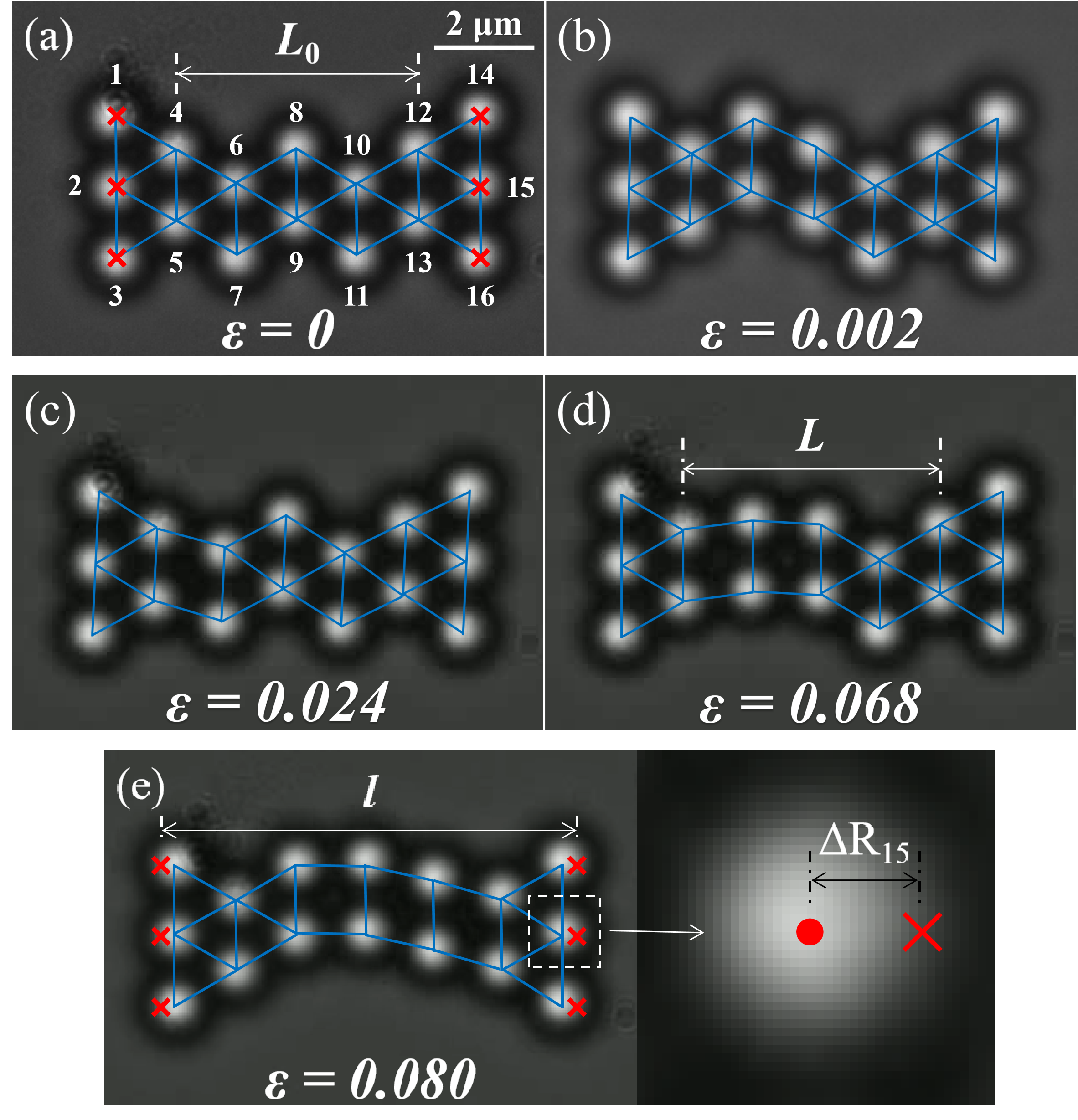}
	\caption{Experimental snapshots of the colloidal cluster at various strains: (a) $\varepsilon=0$, (b) $\varepsilon=0.002$, (c) $\varepsilon=0.024$, (d) $\varepsilon=0.068$, and (e) $\varepsilon=0.080$. Particles (bright blobs) are numbered $1$ to $16$. Blue lines represent interparticle bonds. Red crosses mark optical trap positions. $L_0$ and $L$ correspond to the separation between the second and sixth particle columns at zero and finite strain, respectively. Inset of (e): magnified view of particle $15$ in (e), where $\Delta R_{15}$ indicates the displacement between the particle and its corresponding optical trap.}
	\label{fig:Fig1}
\end{figure}

We sandwich a droplet of sample solution (about $20$ $\mu$L) between two parallel cover glasses ($22\times 40$ mm, Thermo Fisher). The gap between the glass surfaces is slightly larger than the particle size, confining the particle suspension to a quasi-2D plane. The glass sample cell is sealed with UV glue (Norland optical adhesive 65) to avoid solvent evaporation. The sealed sample cell is placed inside a temperature-controlled incubator (UNO, OKOLAB) mounted on the stage of an inverted optical microscope (Ti2U, Nikon).

We image the sample using a $100\times$ oil-immersion objective lens with a numerical aperture of 1.3. A digital camera (acA4112-20um, Basler) captures images at $20$ frames per second with a $4096\times 3000$ pixel resolution. A homemade software is used to locate particle coordinates from the images, achieving a positional precision of about $20$ nm.

Our laser tweezer system (Tweez 305, Aresis) is integrated with a $5$ W infrared laser ($1064$ nm) and an acousto-optical deflector (AOD) operated at 100\,000 Hz. Multiple focused laser spots are generated inside the sample cell through a time-sharing mechanism. Each optical trap can be positioned independently with a spatial resolution of 1 nm. Every laser spot delivers an effective power of 8 mW, which creates a trapping potential of approximately $-14\,k_\text{B}T$ for each trapped particle (see Appendix \ref{appendix_a}).

The ribbon-shaped colloidal cluster comprises $16$ particles. These particles are arranged into two trapezoidal structures at both ends and four rhombic structures in the central region, as shown in Fig.~\ref{fig:Fig1}(a). To assemble this cluster, we first create three laser traps arranged in a vertical array; the positions of these traps are denoted by ${\bm R_i}\equiv \{(X_i,Y_i)\}$ for $i=1,2,3$ with $X_1=X_2=X_3$. The trap spacing ($|Y_1-Y_2|$ and $|Y_2-Y_3|$) is set to be identical to $d$. We then use another laser trap to sequentially move three particles into the trap array sites to form the leftmost particle column. The remaining particles of the cluster are assembled column by column. Columns 2 through 6 each contain two particles, and the rightmost column contains three particles. As the particles are bound by short-range depletion attractions, the cluster can be thought of as part of a 2D close-packed hexagonal particle lattice. 

The particle positions are denoted by $\{\bm{r}_i\}=\{(x_i,y_i)\}$ for $i=1,2,\dots, 16$; see Fig.~\ref{fig:Fig1}(a). The spacing between the leftmost (first) and rightmost (seventh) particle columns is measured by
\begin{equation}
	L'=\frac{1}{3}\left(\sum_{i=14}^{16} x_i-\sum_{i=1}^{3} x_i\right),
	\label{eq_cluster_length}
\end{equation}
whose initial, unstretched value is $L_0'\simeq 7.25\,\mu\text{m}$.

To perform tensile tests on the cluster, we add a second set of optical traps positioned at particles $14$, $15$, and $16$ at the cluster’s right end. The positions of these traps are denoted by $\bm R_i \equiv (X_i, Y_i)$ for $i=14,15,16$, where $X_{14}=X_{15}=X_{16}$, $Y_{14}=Y_1$, $Y_{15}=Y_2$, and $Y_{16}=Y_3$. These positions make the two trap arrays mirror symmetric. The combined configuration of the cluster and optical traps is taken as the ground state for all subsequent tensile measurements. To apply tensile stress, we displace the right-side optical traps at $\bm R_{14}$, $\bm R_{15}$, and $\bm R_{16}$ rightward along the $x$-axis. We neglect shear stress during the tensile test thanks to the mirror symmetry of the two trap arrays.

The separation between the left and right trap arrays is defined by
\begin{equation}
	l=X_{15}-X_2,
	\label{eq_laser_separation}
\end{equation}
whose initial value $l_0=L_0'$ corresponds to zero tensile strain in the cluster. The displacement $\Delta l\equiv l-l_0$ thus serves as the sole control parameter during the tensile test. Figure~\ref{fig:Fig1}(b) shows the snapshot of the stretched colloidal cluster for \(\Delta l=0.06\,\mu\text{m}\). In this case, we find $L'(\Delta l)\simeq 7.27\,\mu\text{m}$ and therefore $L_0'<L'(\Delta l)<l$, confirming that a finite tensile strain emerges in the cluster as a response to $\Delta l$.

We notice that particles within the two end trapezoids, including the first, second, sixth, and seventh particle columns, never rearrange during the tensile test. Particle rearrangement and bond rupture occur only between the second and sixth columns. This observation indicates that deformation concentrates in the central rhombic structures (columns 2 through 6) instead of the trapezoidal ends, likely due to the lower bond energy density in the central region. The trapezoidal segments thus act as wide gripping regions analogous to macroscopic tensile specimens, while the narrower central rhombic zone can be regarded as the specimen gauge section. Following the definition of gauge length in macroscopic tensile tests, we define the tensile strain of the deformable region for a given \(\Delta l\) as
\begin{equation}
	\varepsilon(\Delta l)\equiv \langle L\rangle/L_0 -1,
	\label{eq_strain}
\end{equation}
where
\begin{equation}
	L(\Delta l)=\frac{1}{2}\left(\sum_{i=12}^{13} x_i-\sum_{i=4}^{5} x_i\right)
	\label{eq_cluster_length}
\end{equation}
is the separation between the second and sixth particle columns for an instantaneous cluster configuration. The angular brackets $\langle\dots\rangle$ average over fluctuating particle positions for a given $\Delta l$. \(L_0\equiv \langle L\rangle_{\Delta l=0}\) represents the averaged unstretched value of $L$.

The potential energy of a particle inside the 8-mW laser trap is approximated by a Gaussian potential:
\begin{equation}
	U_\text{opt}(\Delta R_i)=-U_0e^{-2\Delta R_i^2/w^2},
	\label{tweezer_potential}
\end{equation}
where $\Delta R_i\equiv |\Delta \bm{R}_i|=|\bm{R}_i-\bm{r}_i|$ (for $i=1,2,3,14,15,16$) is the distance between laser trap $i$ and the trapped particle, and $w$ is the waist radius of the focused laser beam. The force applied by each laser trap is
\begin{equation}
	\bm{F}(\Delta \bm{R}_i)= \dfrac{4U_0}{w^2}\bm{\Delta R}_i e^{-2\Delta R_i^2/w^2}.
	\label{eq_force}
\end{equation}
When $\Delta R_i$ is small, Eq.~(\ref{eq_force}) can be approximated by
\begin{equation}
	\bm{F}(\Delta \bm{R}_i) \simeq \dfrac{4U_0}{w^2}\bm{\Delta R}_i.
	\label{eq_force_approx}
\end{equation}

The total tensile force is the sum
\begin{equation}
	\tau=\frac{1}{2}\left[\sum_{i=14}^{16}F_x(\Delta \bm{R}_i)-\sum_{i=1}^{3}F_x(\Delta \bm{R}_i)\right],
	\label{eq_total_force}
\end{equation}
where $F_x$ is the (signed) force component along the $x$-axis. $\tau>0$ ($\tau<0$) corresponds to tensile (compressive) force. Due to thermal fluctuations, the instantaneous $\tau$ can be either tensile or compressive. We thus average over the fluctuations and define the (averaged) tensile stress of the cluster as
\begin{equation}
	\sigma(\Delta l)\equiv \langle \tau\rangle/2d.
	\label{eq_stress}
\end{equation}

For some $\Delta l$, the cluster can transition between multiple metastable states. When these states are separated by large energy barriers, the cluster typically explores only a single state during short experimental times. To increase the cluster's propensity to explore the whole energy landscape, we cyclically vary $\Delta l$ to force the cluster to explore all metastable states. Specifically, we rapidly increase $\Delta l$ to \(\Delta l+\delta l \) (e.g.~$\delta l=0.4\,\mu\text{m}$) within $30$ s. This rapid stretching forces the cluster to deform and rearrange into a higher-energy configuration. Then, $\Delta l$ is resumed, after which the cluster is allowed to descend in the energy landscape and re-equilibrate for $5$ s. After equilibration, we sample the fluctuations of $L$ for $30$ s and then repeat this cycle. Within each cycle, the experiment mostly samples one dominant metastable state, while repeated cycles enable the cluster to access different states. This cyclic experiment runs for $90$ min until the sampled distribution of $L$ no longer varies over the sampling time. The cyclically sampled distribution $p(L)$ is assumed to reach equilibrium.

We identify particle configuration $C$ using the coarse-grained connectivity via the adjacency matrix \cite{doi:10.1137/100784424}. The configurational probability $P(C)=N_C/N$ is computed from the number of image frames $N_C$ in which the cluster exhibits configuration $C$, with $N$ the total number of frames.

\section{Theoretical model and numerical simulation}
\label{sec:theory}
\begin{figure}[H]
	\centering
	\includegraphics[width=0.8\linewidth]{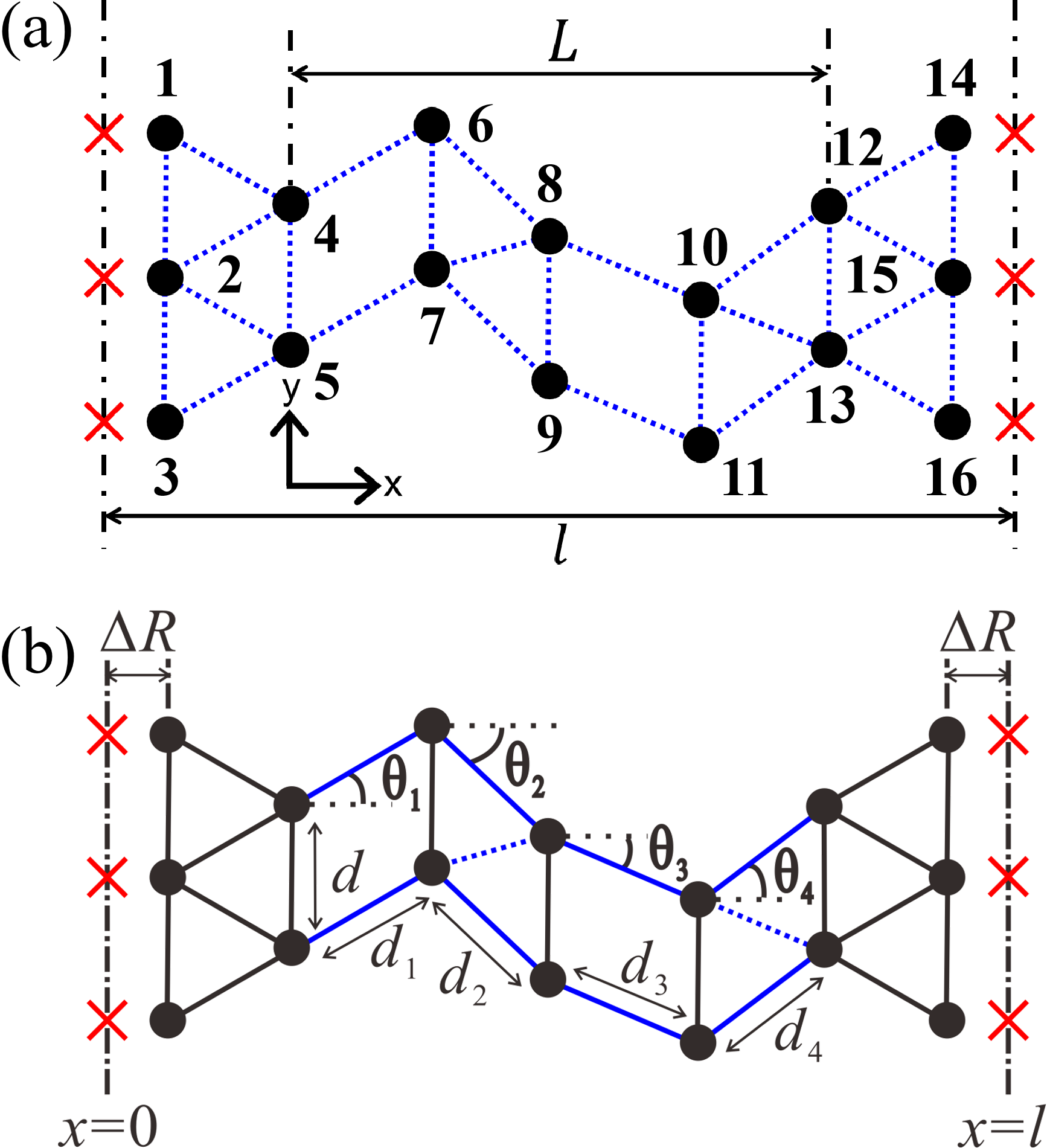}
	\caption{(a) The spring-mass framework with breakable springs (blue dotted lines) and point masses (black circles). The masses (black dots) are labeled from $1$ to $16$, in accordance with Fig.~\ref{fig:Fig1}. Red crosses: optical traps. $l$: distance between the left and right columns of optical traps. $L$: distance between the second and sixth mass columns. (b) The reduced model. Black lines: rigid bonds with length $d$. Blue solid lines: elastic bonds. Blue dotted lines: breakable elastic bonds. $d_{i}$ ($i=1,2,3,4$): lengths of elastic bonds. $\theta_{i}$: bond angles relative to the $x$-axis. $\Delta R$: distance between the optical traps and the trapped particles.}
	\label{fig:Fig_simu_model}
\end{figure}

We model the colloidal cluster as a framework of $16$ point masses connected by breakable elastic springs; see Fig.~\ref{fig:Fig_simu_model}(a). For convenience, we reuse $\{\bm{r}_i\}=\{(x_i,y_i)\}$ ($i=1,2,\dots, 16$) to denote the mass positions. The energy of the spring between a pair of neighboring particles $i$ and $j$ is defined by
\begin{equation}
	U_\text{spr}(\Delta r_{ij})=
	\begin{cases}
		\infty,                 & \Delta r_{ij}<D \\[4pt]
		k_{p}(\Delta r_{ij}-d)^2/2-U_0, & D\leq \Delta r_{ij}\leq d_\text{cut}\\[4pt]
		0,                      & \Delta r_{ij}> d_\text{cut},
	\end{cases}
	\label{sping_simu}
\end{equation}
where $\Delta r_{ij}\equiv|\bm{r}_i-\bm{r}_j|$ is the spring length. $d_{\text{cut}}=1.45$ $\mu$m is the cutoff distance beyond which $U_\text{spr}$ vanishes. $k_p=1.06\times 10^4~k_\text{B}T/\mu \text{m}^2$ and $U_0=6.5$ $k_\text{B}T$ are the spring constant and potential depth, respectively, of a harmonic potential well that approximates the attractive interparticle potential well (see Appendix \ref{appendix_a}). Note that when $d$ and $d_\text{cut}$ both approach $D$ and $k_p$ is very large, Eq.~\ref{sping_simu} effectively describes the interaction between skicky spheres \cite{annurev:/content/journals/10.1146/annurev-conmatphys-031016-025357}.

The total potential energy of the system is the sum:
\begin{equation}
	U = \sum_{i,j}U_\text{spr}(\Delta r_{ij})+\sum_{j}U_\text{opt}(\Delta R_j),
	\label{total_energy}
\end{equation}
where the first sum is over all particle pairs (separated by $\Delta r_{ij}<d_{\text{cut}}$) and the second is over the potential energies of six trapped particles ($j=1,2,3,14,15,$ and $16$).

The full model of Eq.~(\ref{total_energy}) comprises $32$ degrees of freedom for a given $\Delta l$ value, which is employed to perform canonical Monte Carlo simulations (see Appendix \ref{appendix_c}). For analytical analyses, we simplify the model by eliminating less important degrees of freedom based on the following experimental observations: (1) Bonds in the trapezoids never break and are treated as rigid; (2) The particle pair in each column moves as a rigid body, so the intra-column bonds are treated as rigid as well; (3) Particle columns remain approximately perpendicular to the $x$-axis before fracture, so particles in the same column are assigned identical $x_i$ values; (4) Trapped particles by laser tweezers are assigned identical $y_i$ values as the corresponding traps. Under these approximations, only $7$ degrees of freedom remain in the model; see Fig.~\ref{fig:Fig_simu_model}(b). For computational convenience, we use a set of generalized coordinates $q\equiv\{ d_1,d_2,d_3,\theta_1,\theta_2,\theta_3,L\}$, including the lengths ($d_1=\Delta r_{57}$, $d_2=\Delta r_{79}$, and $d_3=\Delta r_{9,11}$) and angles of the non-diagonal springs in the rhombus region, as shown in Fig.~\ref{fig:Fig_simu_model}(b).

We further approximate that the magnitudes of $\{\Delta R_j\}$ are equal to $\Delta R$ for all trapped particles, then the optical potential energy in Eq.~(\ref{total_energy}) reduces to
\begin{equation}
	\sum_{j}U_\text{opt}(\Delta R_j)=-6U_{0}e^{-2\Delta R^2/w^2},
	\label{optical_energy}
\end{equation}
where $\Delta R\equiv (1/2)|l-L-\sqrt{3}\,d|$ is the horizontal displacement between laser traps and trapped particles; see Fig.~\ref{fig:Fig_simu_model}(b). With the total energy defined by Eqs.~(\ref{total_energy}) and (\ref{optical_energy}), any mechanical observable $\langle A\rangle$ ($\langle A\rangle=\varepsilon$ or $\sigma$) can be computed from the ensemble average (see Appendix \ref{appendix_b}):
\begin{equation}
	\langle A\rangle=\frac{1}{2}Z_\text{con}^{-1}\int_M \text{d}q A\, d_1d_2d_3\,e^{-\beta U(q,\Delta l)},
	\label{ensemble_average}
\end{equation}
where $Z_\text{con}$ is the configurational partition function
\begin{equation}
	Z_\text{con}=\frac{1}{2}\int_M \text{d}q d_1d_2d_3\,e^{-\beta U(q,\Delta l)},
	\label{partition_func}
\end{equation}
where the subscript $M$ denotes the accessible configuration space for a given $\Delta l$. The probability of observing configuration $C$ is
\begin{equation}
	P(C)=\frac{1}{2}Z_\text{con}^{-1}\int_{C} \text{d}q d_1d_2d_3\,e^{-\beta U(q,\Delta l)},
	\label{eq_config}
\end{equation}
where the subscript $C$ denotes the sub-configuration space consistent with $C$.

In Section~\ref{sec:results}, we compare Eqs.~(\ref{ensemble_average}) and (\ref{eq_config}) with experimental results. A primary goal of this comparison is to examine whether the tensile response of the colloidal cluster is fully captured by the reduced model. To complement the comparison, and to provide additional information on the cluster under tension, we also present numerical results of the mechanical responses from the simulation using the full model of Eq.~(\ref{total_energy}).

\section{Results and discussion}
\label{sec:results}
\subsection{Probability distribution of cluster length}
\begin{figure}[H]
	\centering
	\includegraphics[width=1\linewidth]{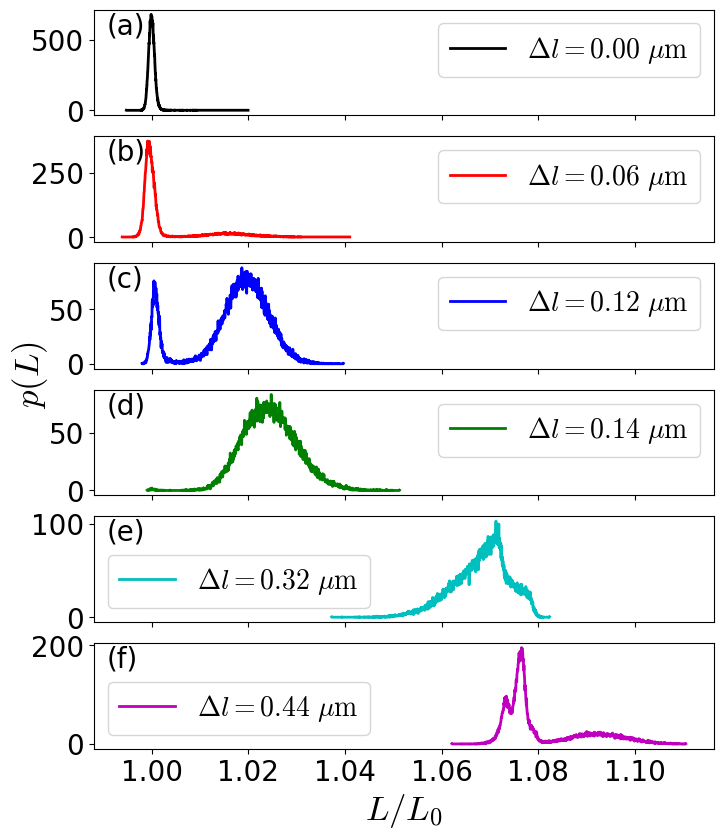}
	\caption{Probability density $p(L)$ of instantaneous $L$ for increasing optical trap separation: (a) $\Delta l=0$, (b) $\Delta l=0.06\,\mathrm{\mu m}$, (c) $\Delta l=0.12\,\mathrm{\mu m}$, (d) $\Delta l=0.14\,\mathrm{\mu m}$, (e) $\Delta l=0.32\,\mathrm{\mu m}$, and (f) $\Delta l=0.44\,\mathrm{\mu m}$.}
	\label{fig:Fig3}
\end{figure}

Particles in the cluster exhibit restricted Brownian motion even when the cluster is clamped by laser traps. The positional fluctuations of the particles in turn induce fluctuations in the deformable length $L$ of the cluster, as characterized by the probability density $p(L)$. Figure~\ref{fig:Fig3} plots $p(L)$ for different $\Delta l$ values. As $\Delta l$ increases, the peak of $p(L)$ shifts toward larger $L$, consistent with the growing tensile strain. For zero tensile strain ($L'_0=l_0=7.25~\mu\text{m}$), the Gaussian-like $p(L)$ resembles the distribution of a one-dimensional (1D) harmonic oscillator in thermal equilibrium; the peak of $p(L)$ corresponds to the mean cluster length $L_0\simeq 4.79~\mu\text{m}$ [Fig.~\ref{fig:Fig3}(a)].

For $\Delta l=0.12~\mu\text{m}$, the secondary $p(L)$ peak at $L/L_0\simeq 1.02$ becomes dominant. This secondary peak has a Gaussian-like shape with a much broader width than the first peak at $L/L_0\simeq 1.0$; the mean cluster length associated with the secondary peak ($L/L_0\simeq 1.020$) is slightly longer than that of the less stretched cluster ($L/L_0\simeq 1.015$) [Fig.~\ref{fig:Fig3}(c)]. This comparison indicates that further deformation increases the probability for the cluster to explore the stretched, longer-length metastable state, whose mean cluster length $L$ also increases.

Further stretching to $\Delta l=0.14~\mu\text{m}$ completely eliminates the probability of observing the unstretched configuration; see Fig.~\ref{fig:Fig3}(d). Interestingly, the resulting Gaussian-like, single-peaked $p(L)$ for $\Delta l \simeq 0.14~\mu\text{m}$ also resembles the fluctuations of a 1D harmonic oscillator, like the unstretched cluster but with a smaller rigidity.

For higher $\Delta l$ values, $p(L)$ exhibits more complex shapes. For $\Delta l\simeq 0.32~\mu\text{m}$, for instance, $p(L)$ is no longer symmetric about its peak at $L/L_0\simeq 1.07$, and a minor sub-peak appears on the right of the main peak; see Fig.~\ref{fig:Fig3}(e). Interestingly, for the strongly stretched cluster at $\Delta l\simeq 0.44~\mu\text{m}$, the overall shape of $p(L)$ looks similar to that of the slightly stretched sample at $\Delta l\simeq 0.06~\mu\text{m}$: a weak secondary peak emerges at $L/L_0\simeq 1.09$, while the primary peak at $L/L_0\simeq 1.07$ splits into two distinct sub-peaks. When $\Delta l>0.48~\mu\text{m}$, the cluster ruptures quickly.

\subsection{Stress-strain curve}
\begin{figure}[H]
	\centering
	\includegraphics[width=0.8\linewidth]{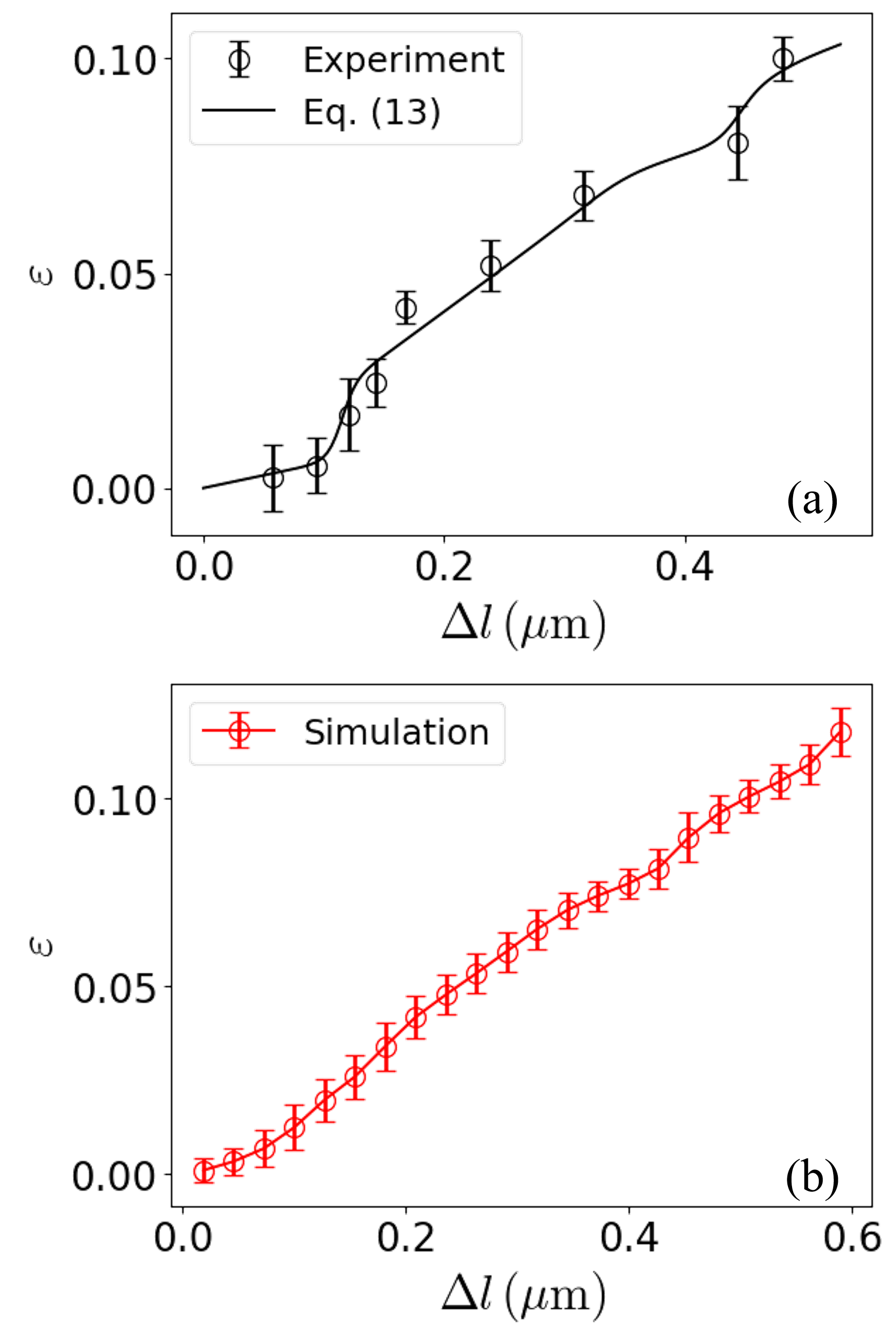}
	\caption{(a) Measured strain $\varepsilon$ vs displacement $\Delta l$ from the experiment. Error bars: standard deviation of $L/L_0-1$ in the experiment. Black curve: theoretical prediction using Eq.~(\ref{ensemble_average}). (b) Measured strain $\varepsilon$ vs displacement $\Delta l$ from the simulation. Error bars: standard deviation of $L/L_0-1$ in the simulation.}
	\label{fig:Fig4}
\end{figure}
The tensile strain of the cluster $\varepsilon$ is directly computed from $p(L)$ via Eq.~(\ref{eq_strain}). Figure~\ref{fig:Fig4}(a) plots $\varepsilon$ for different $\Delta l$ values. The error bar derives from the standard deviation of $p(L)$. Notably, the fracture strain accessible to the experiment reaches $\varepsilon\simeq 0.1$. Remarkably, we observe excellent agreement between the experimental strain and theoretical prediction from the simplified model via Eq.~(\ref{ensemble_average}). The experimental $\varepsilon$ exhibits an overall linear dependence on $\Delta l$, with subtle but noticeable nonlinear regimes for certain $\Delta l$ values. For instance, near $\Delta l\simeq 0.10~\mu\text{m}$ and $0.40~\mu\text{m}$, $\varepsilon$ vs $\Delta l$ exhibits small drops relative to the linear baseline, indicating that tensile deformation is slightly hindered. In other words, the cluster exhibits marginally stronger resistance to tensile force at these $\Delta l$ values, as compared to the overall tensile response. These subtle nonlinear mechanical responses are precisely reproduced by the theoretical curve, suggesting that the simplified model captures essential features of the mechanical response behavior of the colloidal cluster despite the elimination of most degrees of freedom.

Figure~\ref{fig:Fig4}(b) plots $\varepsilon$ from the simulation. The numerical result agrees well with both experimental measurement and theoretical prediction. Notably, the simulation generates more abundant sampling of the cluster configuration, which in turn helps the cluster attain larger values of $\Delta l>0.48~\mu\text{m}$, which are inaccessible to experiment due to frequent fracture.

\begin{figure}[H]
	\centering
	\includegraphics[width=0.8\linewidth]{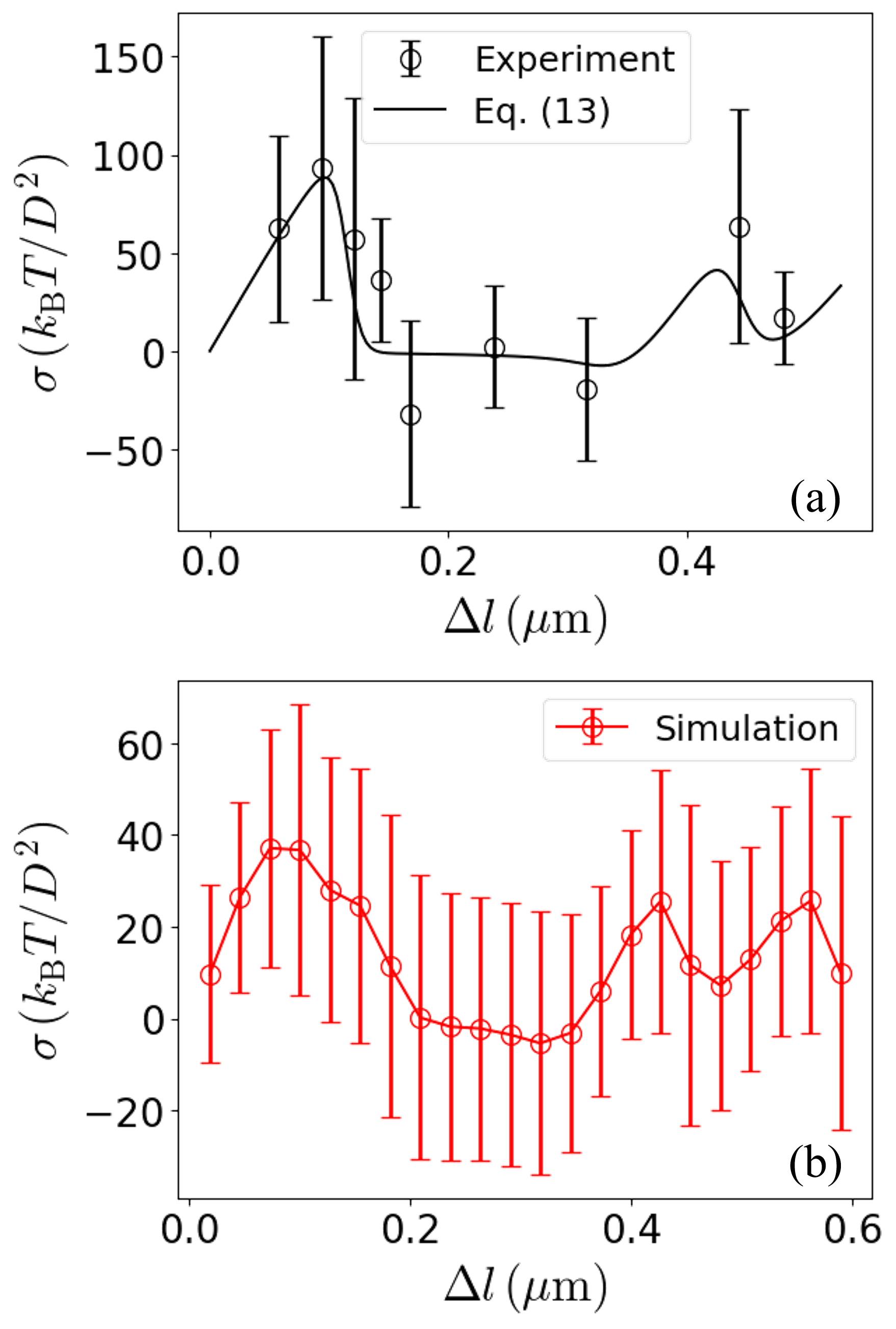}
	\caption{(a) Measured stress $\sigma$ vs displacement $\Delta l$ from the experiment. Error bars: standard deviation of $\tau/2d$ in the experiment. Black curve: theoretical prediction from Eq.~(\ref{eq_stress}). (b) Measured stress $\sigma$ vs displacement $\Delta l$ from the simulation. Error bars: standard deviation of $\tau/2d$ in the simulation.}
	\label{fig:Fig5}
\end{figure}
The tensile stress of the cluster $\sigma$ is computed from positions of the particles (${\bm r}_i$, $i=1,2,3,14,15,$ and $16$) and laser traps (${\bm R}_i$, $i=1,2,3,14,15,$ and $16$) via Eq.~(\ref{eq_stress}). Figure~\ref{fig:Fig5}(a) plots $\sigma$ for various $\Delta l$ values. The error bars originate from fluctuations in particle positions for each $\Delta l$, whereas trap positions are fixed to predetermined values. Upon stretching, $\sigma$ increases rapidly to a maximum of $\sigma\simeq 90~k_BT/D^2$ at $\Delta l\simeq 0.10~\mu\text{m}$, then descends to approximately zero on the negative side; according to Eq.~(\ref{eq_stress}), negative $\sigma$ signals compressive rather than tensile stress. This observation suggests that, at intermediate $\Delta l$ (e.g., $0.18~\mu\text{m}<\Delta l<0.36~\mu\text{m}$), the weak stress resembles soft-mode-like deformation, which is strongly susceptible to external force. A secondary peak in $\sigma$ appears upon further stretching to $\Delta l\simeq 0.40~\mu\text{m}$; beyond this, $\sigma$ decays to nearly zero again. Notably, the two peaks in the $\sigma$ vs $\Delta l$ curve occur at the same $\Delta l$ values where $\varepsilon$ exhibits slight drops, respectively [Fig.~\ref{fig:Fig4}(a)]. We thus conclude that both the $\sigma$ peaks and $\varepsilon$ drops are manifestations of the cluster's increased resistance to tensile deformation.

Remarkably, our simplified cluster model quantitatively reproduces the two stress peaks observed experimentally; see Fig.~\ref{fig:Fig5}(a). The model further predicts a soft-mode-like behavior with nearly zero compressive stress. The predicted compressive strength, however, is smaller than the experimental values, which we suspect may result from either insufficient experimental sampling or oversimplification of the reduced cluster model.

Figure~\ref{fig:Fig5}(b) plots $\sigma$ vs $\Delta l$ from the simulation. The numerical result qualitatively reproduces both the experimental and theoretical results. Specifically, the numerical result displays two stress peaks and a soft-mode-like region with compressive stress at approximately the same $\Delta l$ values as the experiment and theoretical prediction. The extended range of $\Delta l$ accessible to the simulation even reveals a third stress peak at $\Delta l\simeq 0.57~\mu\text{m}$, which is difficult to observe experimentally due to frequent fracture near this range of $\Delta l$. The excellent agreement among experiment, theory, and simulation suggests that our theoretical model, including its simplified approximation, adequately captures the rich mechanical response of the colloidal cluster.

\begin{figure}[H]
	\centering
	\includegraphics[width=0.9\linewidth]{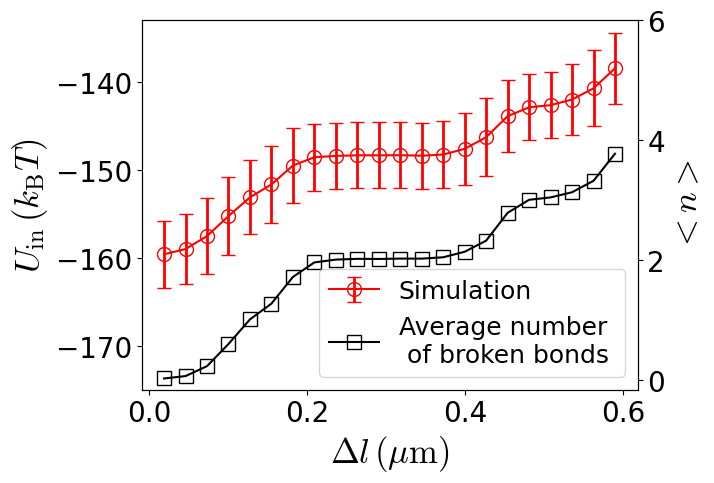}
	\caption{Measured internal energy $U_\text{in}$ and average number of broken bonds $\langle n\rangle $ as a function of displacement $\Delta l$ from the simulation. Error bars: standard deviation of $\sum_{i,j}U_\text{spr}(\Delta r_{ij})$ and $n$.}
	\label{fig:Fig6}
\end{figure}

To further characterize the soft-mode-like response regime, we extract the cluster’s internal energy $U_\text{in}\equiv \bigl\langle \sum_{i,j}U_\text{spr}(\Delta r_{ij})\bigr\rangle$ as a function of $\Delta l$ from simulation data; see Fig.~\ref{fig:Fig6}. $U_\text{in}$ increases monotonically with $\Delta l$ and forms a plateau for $0.20\,\mu\text{m}\lesssim \Delta l\lesssim 0.38\,\mu\text{m}$. This plateau region indicates that $U_\text{in}$ stays nearly constant during the soft-mode deformation (in simulation). The plateau range from simulations agrees well with the experimentally observed soft-mode regime $0.18\,\mu\text{m}\lesssim \Delta l\lesssim 0.36\,\mu\text{m}$, confirming the fidelity of our full model in simulation. Interestingly, $U_\text{in}$ vs $\Delta l$ exhibits a qualitatively identical trend as the average number of broken bonds $\langle n\rangle$, as shown in Fig.~\ref{fig:Fig6}. No bond is broken within the soft-mode deformation regime.

\begin{figure}[H]
	\centering
	\includegraphics[width=0.8\linewidth]{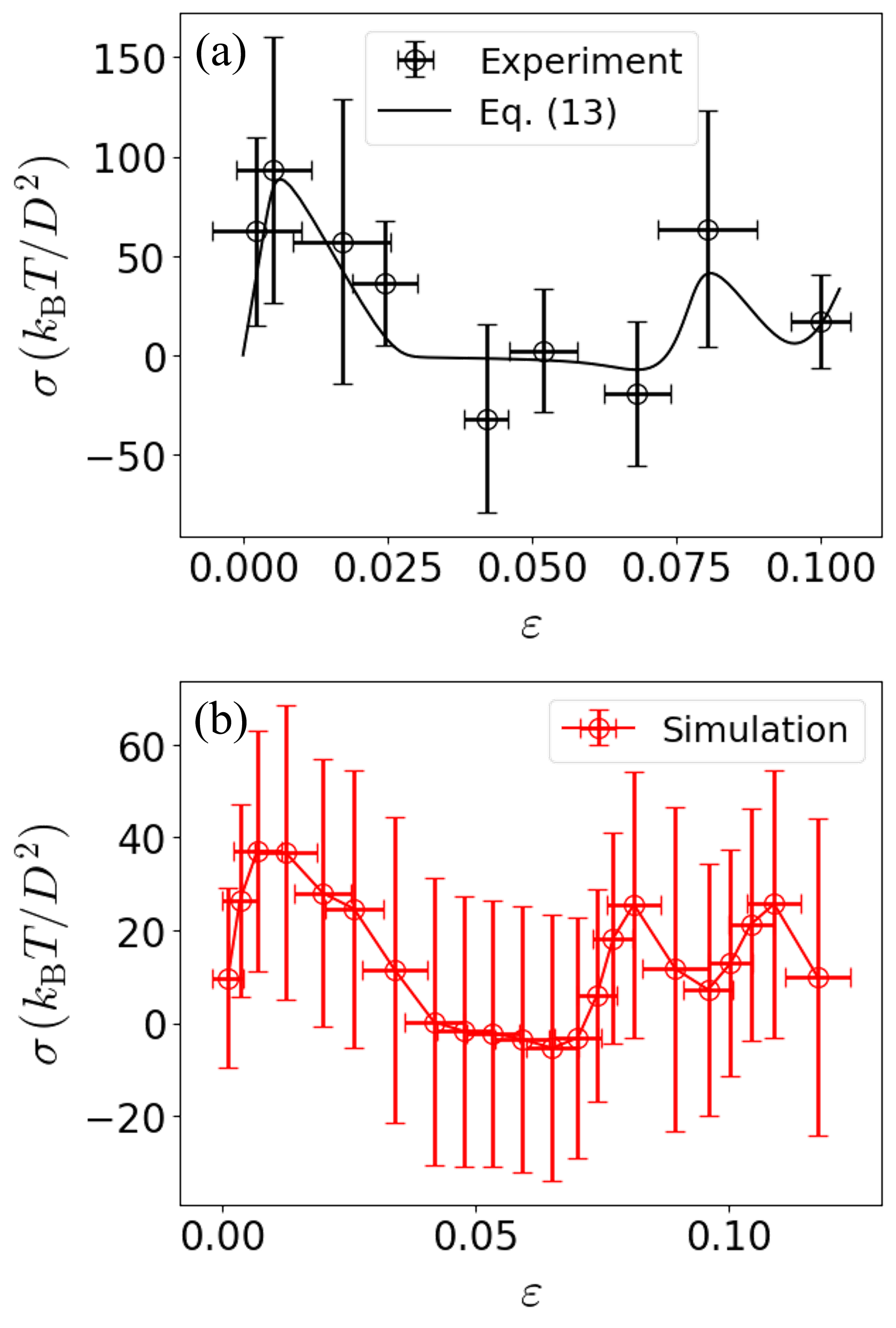}
	\caption{(a) Measured $\sigma$ vs $\varepsilon$ from the experiment. Black curve: theoretical prediction from Eqs.~(\ref{eq_strain}) and (\ref{eq_stress}). Error bars are identical to those in Figs.~\ref{fig:Fig4}(a) and \ref{fig:Fig5}(a). (b) Measured $\sigma$ vs $\varepsilon$ from the simulation (red circles). Error bars are identical to those in Figs.~\ref{fig:Fig4}(b) and \ref{fig:Fig5}(b).}
	\label{fig:Fig7}
\end{figure}

To fully characterize the mechanical response of the colloidal cluster, we replot $\sigma$ vs $\varepsilon$ for both experimental result and theoretical prediction in Fig.~\ref{fig:Fig7}(a). At small strains $\varepsilon \lesssim 0.01$, $\sigma$ increases monotonically with $\varepsilon$, indicating elastic deformation as confirmed by experimental imaging where all interparticle bonds remain intact [Fig.~\ref{fig:Fig1}(a)]. For $0.01 \lesssim \varepsilon \lesssim 0.03$, $\sigma$ decreases from its maximum value, accompanied by the sequential rupture of two bonds [Fig.~\ref{fig:Fig1}(c,d)]---behavior characteristic of plastic deformation. In the interval $0.03 \lesssim \varepsilon \lesssim 0.07$, $\sigma$ remains near zero, implying negligible work is required for deformation, a hallmark of a soft mode. Beyond $\varepsilon \simeq 0.07$, $\sigma$ first increases then decreases, during which one bond ruptures [Fig.~\ref{fig:Fig1}(e)]. This second elastic- and plastic-like response behavior appears similar to that observed during initial stretching for $\varepsilon \lesssim 0.03$.

As a comparison, Fig.~\ref{fig:Fig7}(b) plots the stress-strain curve from the simulation data. The result precisely reproduces the full sequence of mechanical regimes observed experimentally, including the initial elastic response, plastic deformation, soft mode, as well as the second and even the third elastic-plastic regimes. Two bonds are observed to break during the first plastic regime and one bond breaks during the second plastic regime, also in agreement with the experiment. Combining experimental and simulation results, we conclude that cluster rupture requires breaking $3$ to $4$ diagonal bonds within the rhombus structures in the middle of the cluster ribbon.

\subsection{Probability distribution of configurations}
\begin{figure}[h]
	\centering
	\includegraphics[width=1\linewidth]{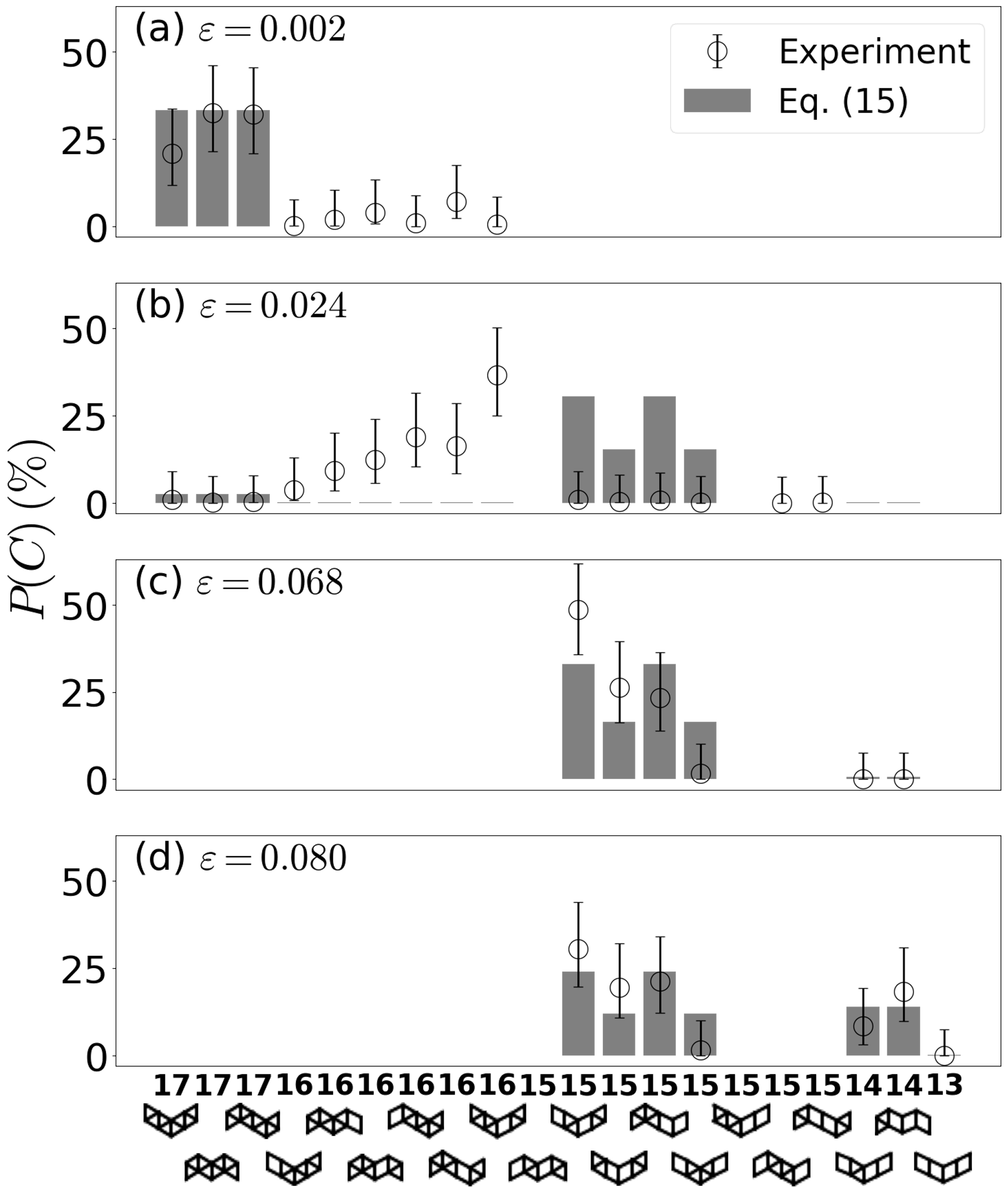}
	\caption{Probability density of distinct cluster configurations, $P(C)$, measured from the experiment (open black circles) for (a) $\varepsilon = 0.002$, (b) $0.024$, (c) $0.068$, and (d) $0.080$. Error bars: 95\% confidence intervals. Grey bars: theoretical predictions from Eq.~(\ref{eq_config}). The connectivity diagrams at the bottom denote distinct cluster configurations $C$ with numbers indicating the total number of interparticle bonds.}
	\label{fig:Fig8}
\end{figure}

\begin{figure}[h]
	\centering
	\includegraphics[width=1\linewidth]{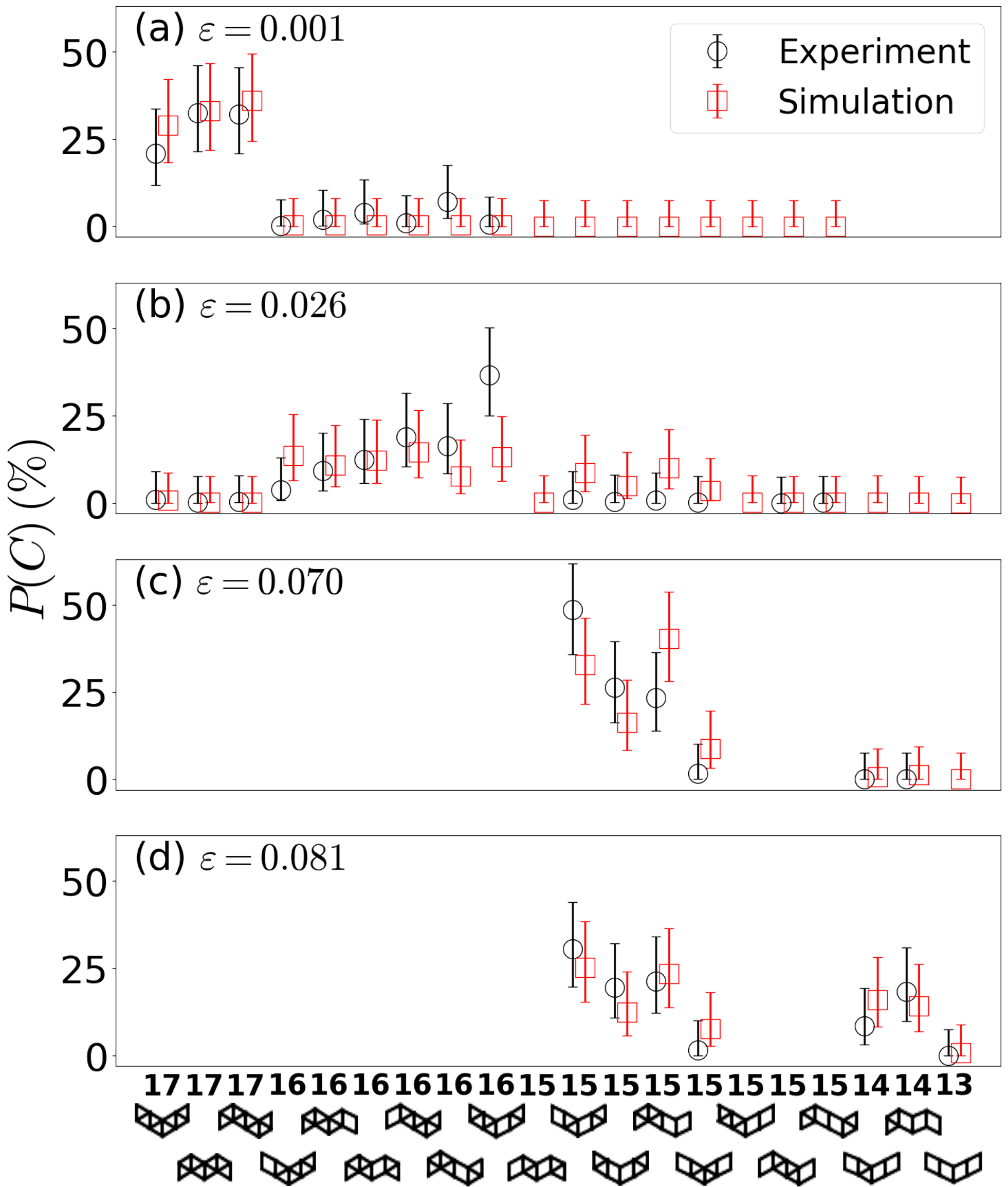}
	\caption{Probability of distinct cluster configurations $P(C)$ measured from the simulation (red squares) for (a) $\varepsilon = 0.001$, (b) $0.026$, (c) $0.070$, and (d) $0.081$.}
	\label{fig:Fig9}
\end{figure}
Finally, we examine the probability distribution of particle configurations $P(C)$ during the tensile test. Figure~\ref{fig:Fig8} shows $P(C)$ for various $\varepsilon$ values, along with the theoretical prediction from the simplified model via Eq.~(\ref{eq_config}). Black schematic diagrams below the figure panels represent the topologies of corresponding configurations (bonds from the second to sixth columns), with the total number of bonds labeled above these diagrams. Within the first elastic regime ($\varepsilon\simeq 0.002$), $C$ predominantly exhibits the ground-state configuration with $17$ bonds; see Fig.~\ref{fig:Fig8}(a). The probability of observing an excited state is nearly zero. Both experimental and simulation results are in excellent agreement with model prediction; see Fig.~\ref{fig:Fig9}(a).

For $\varepsilon \simeq 0.024$, where the cluster is stretched into the first plastic regime, the experimental $P(C)$ deviates significantly from the theoretical prediction. Notably, configurations with $16$ bonds dominate in experimentally observed configurations, whereas the model predicts a prevalence of $15$-bond configurations. The simulation supports the experimental finding [Fig.~\ref{fig:Fig9}(b)], leading us to attribute the discrepancy between experiment and theory to the oversimplifications made in the reduced model. We observe that, for certain $\Delta l$, the particle columns are not perfectly aligned perpendicular to the $x$-axis. For example, in the first plastic deformation regime ($0.01\lesssim\varepsilon \lesssim 0.03$), immediately after the first bond breaks, the first and seventh particle columns tilt slightly clockwise and are no longer perpendicular to the $x$-axis; see Fig.~\ref{fig:Fig1}(b,c). Upon further stretching, when the second bond breaks, these columns return to an orientation perpendicular to the horizontal axis ; see Fig.~\ref{fig:Fig1}(d). This behavior suggests that, within the first plastic regime, the optical traps impose not only tensile strain but also shear strain on the cluster. The simplified model assumes that all particle columns remain vertical and therefore cannot capture shear deformations, which necessarily involve rotations of the particle columns. In principle, the accuracy of the model could be improved by introducing additional degrees of freedom to account for these rotational fluctuations. However, incorporating such detail would substantially increase the computational cost for the partition function and is therefore left for future work.

In the soft-mode and second elastic regimes, we observe excellent agreement between experimental results [Figs.~\ref{fig:Fig8}(c) and (d)], theoretical predictions, and simulations [Figs.~\ref{fig:Fig9}(c) and (d)]. These comparisons further confirm that the simplified cluster model can accurately describe the response behavior in the elastic and soft-mode regimes.

\section{Summary}
\label{sec:summary}
We prepared a quasi-2D ribbon-shaped colloidal cluster of $16$ particles and measured its tensile response to a pair of optical trap arrays and the accompanying particle configurations. Due to the small particle number, the cluster exhibited strong thermal fluctuations in its instantaneous strain, stress, and configurations throughout the tensile testing. The fluctuation-averaged stress--strain curve revealed a sequence of mechanical response regimes: elastic-like deformation, plastic-like deformation, soft-mode response, and a second elastic-plastic transition, before the cluster finally broke at a fracture strain near 10\%. This ductile-like behavior originates from the soft-mode deformation over a large strain range approximately between 3\% and 7\%.

Our sample differs qualitatively from the stress-free equilibrium colloidal clusters studied previously \cite{PhysRevLett.114.228301}. External stress induces anisotropy in the cluster configurations and distinct responses in interparticle bonds with different orientations. As a result, full information on the strain-dependent interparticle energy is required to compute the statistical ensemble. Therefore, we constructed a spring-network model incorporating breakable interparticle bonds. Guided by experimental observations, we reduced the $32$ degrees of freedom of the full model down to $7$. The full model enables precise canonical Monte Carlo simulations that closely reproduce the experimental observations, while the simplified model permits analytical predictions of the deformation and response behaviors and configuration distributions across the full strain range using the unstressed ground-state configuration. We found excellent agreement in the cluster's stress-strain curve among experiment, simulation, and theory, confirming that our minimal model captures the essential physics underpinning the cluster's tensile response behavior.

In addition, we analyzed the evolution of particle configurations of the cluster under tension. The minimal model accurately reproduced experimental distributions in the elastic and soft-mode deformation regimes. Discrepancies between experiment and theory emerge exclusively within the plastic deformation regime, likely due to the oversimplification of the reduced model. This discrepancy implies a limitation of the approximated model and points to future theoretical refinement.

This work provides a versatile experimental platform for studying the mechanics of colloidal clusters. The proposed theoretical framework is scalable; the model can be readily extended to other deformation modes, such as shear and bending. Future efforts may further explore how cluster width regulates the multistage mechanical responses observed in our narrow, two-particle-wide sample. In the limit of large cluster widths, these clusters are expected to transition toward brittle fracture, mirroring the failure behavior of single crystals. This information may yield insights into our fundamental understanding of cluster mechanics and offer practical guidance for tailoring material properties, including the failure behavior of colloidal assemblies.

\section*{Author contributions}
YY and XM designed the research. YY performed the experiments and derived the theory. JK and YL performed the simulations. All analyzed the data and wrote the paper.   

\appendix
\section{Optical trap potential and depletion attractions}
\label{appendix_a}
\renewcommand{\thefigure}{A\arabic{figure}}
\setcounter{figure}{0}
To calculate the optical potential $U_\text{opt}$, we record 20000 images of a trapped particle and obtain the probability density of its displacement $p(\Delta x)$ along the $x$-axis. $p(\Delta x)$ is related to the trapping potential by
\begin{equation}
	p(\Delta x)=Ae^{-U_\text{opt}(\Delta x)/k_BT},
	\label{p_dx}
\end{equation}
where $A$ is a normalization factor. We plot $p(\Delta x)$ in Fig.~\ref{fig:Fig A1}. The potential energy is thus computed by $U_\text{opt}(\Delta x)=-k_BT\ln[p(\Delta x)/A]$. The best fit of $U_\text{opt}(\Delta x)$ to Eq.~(\ref{tweezer_potential}) yields $U_0=13.66\pm 0.01k_\text{B}T$ and $w=0.192\pm 0.002\mu\text{m}$.
\begin{figure}[h]
	\centering\includegraphics[width=0.9\linewidth]{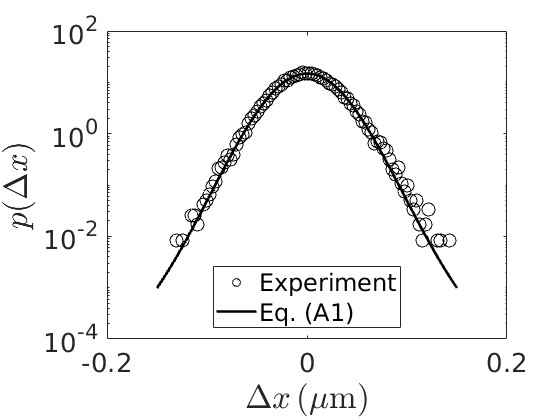}
	\caption{Probability distribution of displacement between the particle and the optical trap.}
	\label{fig:Fig A1}
\end{figure}

The interparticle potential is measured by calibrated optical traps with known force constant using the similar method described in Refs. \cite{polin2008autocalibrated,rudolf2024temporal}. Briefly, two particles are initially trapped with a separation of $r=1.55\,\mu\text{m}$, a distance sufficiently long so that they do not interact with one another. The optical traps are then periodically switched on for 0.1 s and off for 5 s. The probability density of interparticle distance $\psi(r)$ is extracted during the off-state period of each cycle. The interparticle pair potential is calculated by $U(r)=-k_\text{B}T\ln[\psi(r)/r]-u_0$, where the constant $u_0$ is determined such that $U(r)$ vanishes at large separations ($r>1.45\,\mu\text{m}$).

\begin{figure}[h]
	\centering\includegraphics[width=0.9\linewidth]{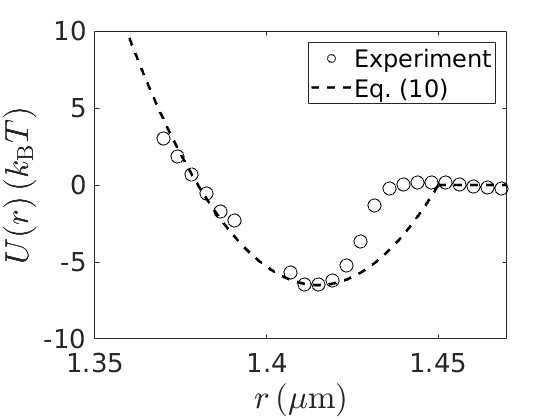}
	\caption{Measured interparticle pair potential $U(r)$. Black curve: interparticle potential used in the sping-mass framework model.}
	\label{fig:Fig A2}
\end{figure}

\section{Derivation of the configurational partition function}
\label{appendix_b}
We treat the two trapezoids and three middle particle columns as rigid bodies. Their center-of-mass positions and momenta are denoted as $\bm{q}'_1, \dots, \bm{q}'_5$ and $\bm{p}_1, \dots, \bm{p}_5$, respectively. Neglecting rotational motion, the Hamiltonian for a given $l$ simplifies to
\begin{equation}
	H=\sum_{i=1}^{5}\frac{\bm{p}_i^2}{2m_i}+U,
	\label{hamiltonian}
\end{equation}
where $m_1=m_5=5m$ are the masses of the trapezoids, and $m_2=m_3=m_4=2m$ are the masses of the particle columns, with $m$ the mass of a single particle. Since $\bm{q}'_1$ and $\bm{q}'_5$ are dependent, the generalized coordinates and momenta can be rewritten as $q'\equiv\{\bm{q}'_1, \dots, \bm{q}'_4\}$ and $p\equiv \{\bm{p}_1, \dots, \bm{p}_4\}$, respectively. The canonical distribution reads
\begin{equation}
	\rho(q',p,l)=\frac{1}{Z}e^{-\beta H(q',p,l)},
	\label{distribution}
\end{equation}
with the partition function $Z$ defined as
\begin{equation}
	\begin{split}
		Z=\iint e^{-\beta H(q',p,l)}\mathrm{d}q'\mathrm{d}p=Z_p\int_M e^{-\beta U}\mathrm{d}q',
	\end{split}
\end{equation}
where the subscript $M$ denotes the accessible configuration space for a given $l$. 

The momentum-space partition function $Z_p$ cancels out in $\rho$, so the distribution is determined by the configuration space:
\begin{equation}
	\rho(q',l)=Z_\text{con}^{-1}e^{-\beta U},
	\label{config_distribution}
\end{equation}
where the configurational partition function reads
\begin{equation}
	Z_\text{con}=\int_M e^{-\beta U}\mathrm{d}q'.
	\label{config_partition}
\end{equation}

To evaluate $Z_\text{con}$, we introduce a new set of generalized coordinates $q\equiv \{d_1,d_2,d_3,\theta_1,\theta_2,\theta_3,L\}$. The Jacobian determinant of the coordinate transformation from $q$ to $q'$ is $-d_1d_2d_3/2$, and Eq.~(\ref{config_distribution}) can be rewritten as
\begin{equation}
	\rho(q,l)=Z_\text{con}^{-1}\frac{d_1d_2d_3e^{-\beta U}}{2}.
	\label{general_distribution}
\end{equation}
To determine the integration domain $M$ for Eq.~(\ref{config_partition}), we consider the constraints among $\{d_i\}$ and $\{\theta_i\}$ ($i=1,2,3,4$):
\begin{equation}
	\sum_{i=1}^{4}d_{i}\cos\theta_i=L,
	\label{cons1}
\end{equation}
and
\begin{equation}
	\sum_{i=1}^{4}d_i\sin\theta_i=0.
	\label{cons2}
\end{equation}
These constraints allow us to express $\theta_4$ and $d_4$ as
\begin{equation}
	\tan\theta_4=\frac{\sum_{i=1}^{3}d_i\sin\theta_i}{\sum_{i=1}^{3}d_i\cos\theta_i-L},
	\label{tan_theta}
\end{equation}
and
\begin{equation}
	d_4^2=\left(\sum_{i=1}^{3}d_i\sin\theta_i\right)^2+\left(\sum_{i=1}^{3}d_i\cos\theta_i-L\right)^2.
	\label{d4}
\end{equation}

Moreover, the angles $\{\theta_i\}$ should be bounded by
\begin{equation}
	|\sin\theta_i|\leq \frac{d_\text{cut}^2+d^2-D^2}{2dd_\text{cut}},
	\label{cons3}
\end{equation}
whereas $L$ should be bounded by
\begin{equation}
	4\sqrt{D^2-d^2/4}\leq L\leq 4d_\text{cut}.
	\label{cons4}
\end{equation}

Thus, the integration domain $M$ in $q$ space should satisfy $D\leq d_{i}\leq d_\text{cut}$ and $D\leq \Delta r_{ij}$, together with Eqs.~(\ref{cons3}) and (\ref{cons4}).

\section{Monte Carlo simulation}
\label{appendix_c}
\renewcommand{\thefigure}{C\arabic{figure}}
\setcounter{figure}{0}
Using Eq.~(\ref{total_energy}) with the full 32 degrees of freedom, we perform canonical Monte Carlo simulations of an NVT ensemble using the Metropolis algorithm. The separation between the two optical trap arrays is adjusted from $l=7.37$ $\mu$m to $7.94$ $\mu$m in increments of $0.03$ $\mu$m to match the experimental conditions.

\begin{figure}[h]
	\centering\includegraphics[width=0.7\linewidth]{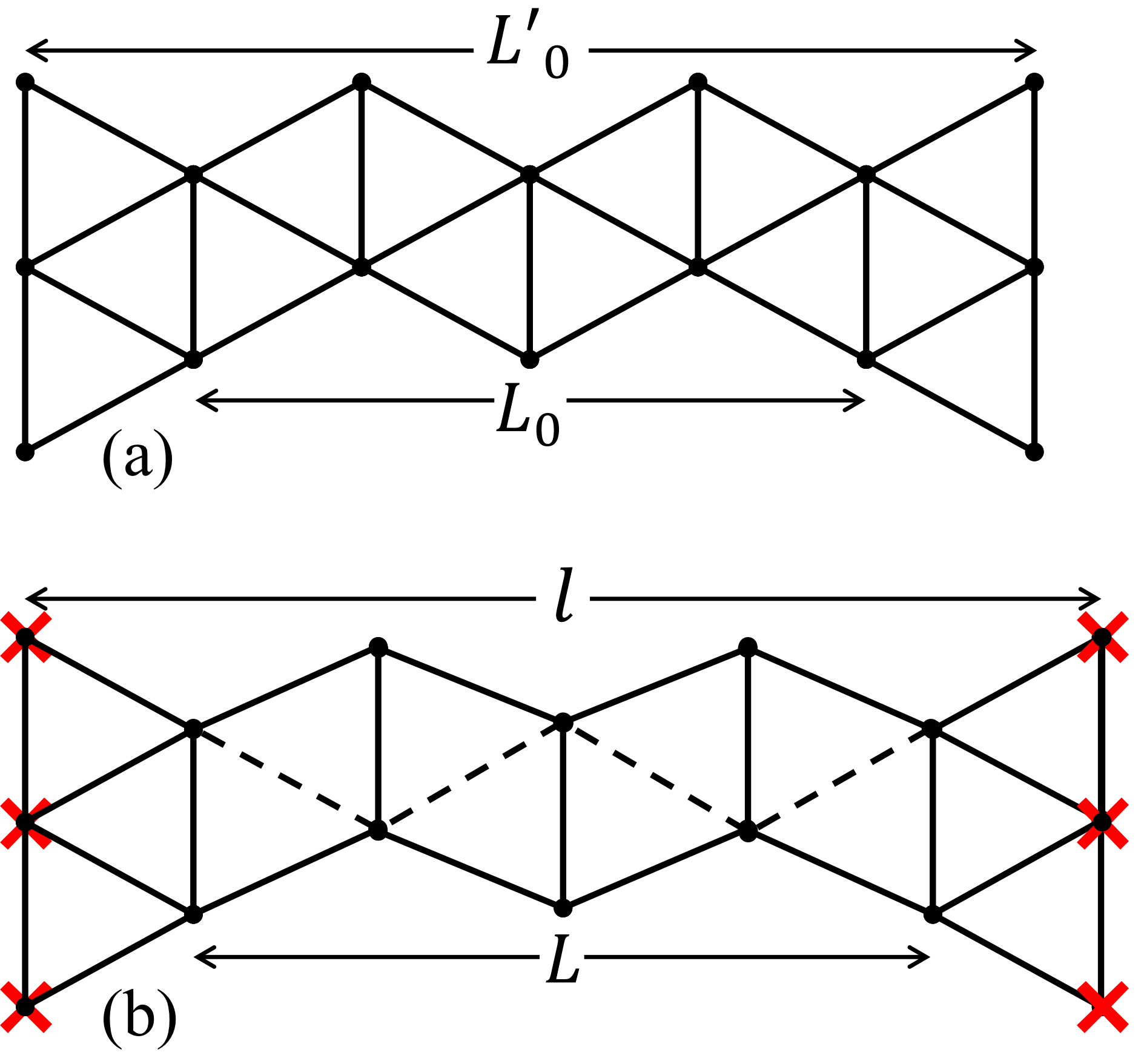}
	\caption{(a) Ground-state configuration. (b) The framework is stretched by increasing $L$ such that the first and seventh mass columns both reach the optical traps (red crosses). During this process, only four diagonal bonds (black dashed lines) are elongated.}
	\label{fig:Fig_stretch_relax}
\end{figure}

For each $l$, we first prepare a spring-mass frame in the ground-state configuration as illustrated in Fig.~\ref{fig:Fig_stretch_relax}(a). In this ground state, all bond lengths are identical to $d$. Next, we increase the separation between the second and sixth particle columns $L$ until the separation between the first and seventh particle column $L'=l$; see Fig.~\ref{fig:Fig_stretch_relax}(b). During this process, all bond lengths are maintained constant ($d$) except for the diagonal bonds (dashed lines), which are stretched to longer values. Subsequently, the $6$ optical traps are positioned at the locations of particles $1$, $2$, $3$, $14$, $15$, and $16$. Finally, the system is allowed to thermally equilibrate, and configuration fluctuations are recorded after the system reaches equilibrium.

\section*{Acknowledgments}
YY and XM thanks the National Natural Science Foundation of China (Grant No. 12274195), National Key Research and Development Program of China (Grant No. 2022YFA1405002), Department of Science and Technology of Guangdong Province (Grant No. 2021QN02C382). JK and YL thank the National Natural Science Foundation of China (Grant No. 12275137) and thank Jeff Chen, Baohui Li, Zheng Wang and Nan Qi for discussions and helps on simulations.

\bibliography{cluster}
\end{document}